\documentclass[11pt]{article}
\usepackage[margin=1in]{geometry}
\usepackage{graphicx}

\def\beq{\begin{equation}}
\def\eeq#1{\label{#1}\end{equation}}
\def\eeqn{\end{equation}}

\def\beqa{\begin{eqnarray}}
\def\eeqa#1{\label{#1}\end{eqnarray}}
\def\eeqan{\end{eqnarray}}

\let\bar=\overbar

\def\Dslash{\not{\hbox{\kern-4pt $D$}}}
\def\dslash{\not{\hbox{\kern-2pt $\del$}}}

\def\msb{{\bar{\ssstyle M \kern -1pt S}}}

\usepackage{mathtools}
\usepackage[hidelinks]{hyperref}
\usepackage{amsmath}
\usepackage{multicol}
\usepackage{amsmath}
\usepackage[normalem]{ulem}
\usepackage{subcaption}
\usepackage{caption}
\usepackage{wrapfig}
\usepackage[T1]{fontenc}
\usepackage[utf8]{inputenc}
\usepackage{authblk}
\usepackage{setspace}
\usepackage[symbol]{footmisc}

\usepackage{epsfig}
\usepackage[utf8]{inputenc} 
\usepackage[T1]{fontenc}
\usepackage{mathptmx}
\usepackage{url}
\usepackage[capitalise]{cleveref}
\newcommand\tab[1][1cm]{\hspace*{#1}}

\def\correspondingauthor{\footnote{Corresponding author\\ \tab[0.5cm] Email address: mounia.laassiri@gmail.com (M. Laassiri)}}
\emergencystretch=\maxdimen
\def\Title#1{\begin{center} {\Large {\bf #1} } \end{center}}
\def\Author#1{\begin{center} {\normalsize {\sc #1} } \end{center}}
\def\Institution#1{\begin{center} {\normalsize {\it #1} } \end{center}}
\def\Abstract#1{\noindent {\normalsize {\bf Abstract:} {\normalfont #1}}}
\def\Conference{\vspace{4mm}\begin{raggedright} {\normalsize {\it Talk presented by M. Laassiri at the 2019 Meeting of the Division of Particles and Fields of the American Physical Society (DPF2019), July 29--August 2, 2019, Northeastern University, Boston, C1907293.} } \end{raggedright}\vspace{4mm}}

\begin{document}

%
%
\Title{The African School of Fundamental Physics and Applications (ASP)}

\Author{Kétévi Adiklè Assamagan$^1$, Mounia Laassiri$^{1,2}$ \correspondingauthor{}}
\Institution{$^1$ Brookhaven National Laboratory, Department of Physics, Upton, New York, 11983, USA} 
\vspace*{-0.8cm}
\Institution{$^2$ ESMaR, Faculty of Sciences, Mohammed V University, 1014 RP, Rabat, Morocco}

\Abstract{The African School of Fundamental Physics and Applications is a biennial school in Africa. It is based on the observation that fundamental physics provides excellent motivation for students of science. The aim of the school is to build capacity to harvest, interpret, and exploit the results of current and future physics experiments and to increase proficiency in related applications. The participating students are selected from all over Africa. The school also offers a workshop to train high school teachers, outreach to motivate high school pupils and a physics conference to support a broader participation of African research faculties. Support for the school comes from institutes in Africa, Europe, USA and Asia. In this paper, we present the school and discuss strategies to make the school sustainable.}

\Conference

%
%

\section{Introduction}

Schools of fundamental physics and applications ~\cite{ASP} took place in Stellenbosch, South Africa, on August 1-21, 2010 ~\cite{ASP2010}, in Kumasi, Ghana, on July 15-August 8, 2012 ~\cite{ASP2012}, in Dakar, Senegal on August 3-23, 2014 ~\cite{ASP2014}, in Kigali, Rwanda on August 1–19, 2016 ~\cite{ASP2016} and Windhoek, Namibia on June 24-July 14, 2018 ~\cite{ASP2018}. The next edition of the school is planned in 2020 in Morocco. 

The basic objective is to help improve the quality of higher education in Africa and increase the number of African students acquiring higher education. This is achieved through an outreach effort, an increased awareness of the potential of high-quality training offered by large scale experiments in context of various scientific disciplines, and a system of networking on the international scale. There is a strong alignment between the mission and the vision of African governments and policy makers on education and capacity building and their programs with the goals of the ASP. The ASP is committed to include African governments in the planning, in order to take advantage of aspects such as consolidating agreements and their goals, building on synergy with other programs, improving the sustainability and impact of capacity development and improving the measurement and visibility of the impact. By working with African governments and policy makers on education, ASP seeks to promote a culture of science that creates an attractive environment for African student alumni, thus encouraging their retention within Africa. ASP promotes sustainable scientific development in Africa by building a network between African and international researchers for increased collaborative research and shared expertise.

The schools are based on a close interplay between theoretical, experimental, applied physics, and Grid computing. They cover a wide range of topics: particle physics, particle detectors, astro-particle physics and cosmology, computing, accelerator technologies, medical physics, condensed matter, light sources and their applications ~\cite{connell2018african}. Scientists from Africa, Europe, Asia and the USA are invited to prepare and deliver lectures according to the proposed topics considering the diverse levels of the students. The duration of the school allows for networking interactions among students and between students and lecturers. The schools are funded by institutes in Africa, Asia, Europe and the USA. ASP serves at least two purposes: the organization of the school creates understanding of the many challenges and provides a template for solving them, and secondly it provides opportunity to prepare the students to find practical answers to many issues that they face.

\section{Venue and scope}

\begin{figure}[!htp]
\centering
\includegraphics[height=2.8in]{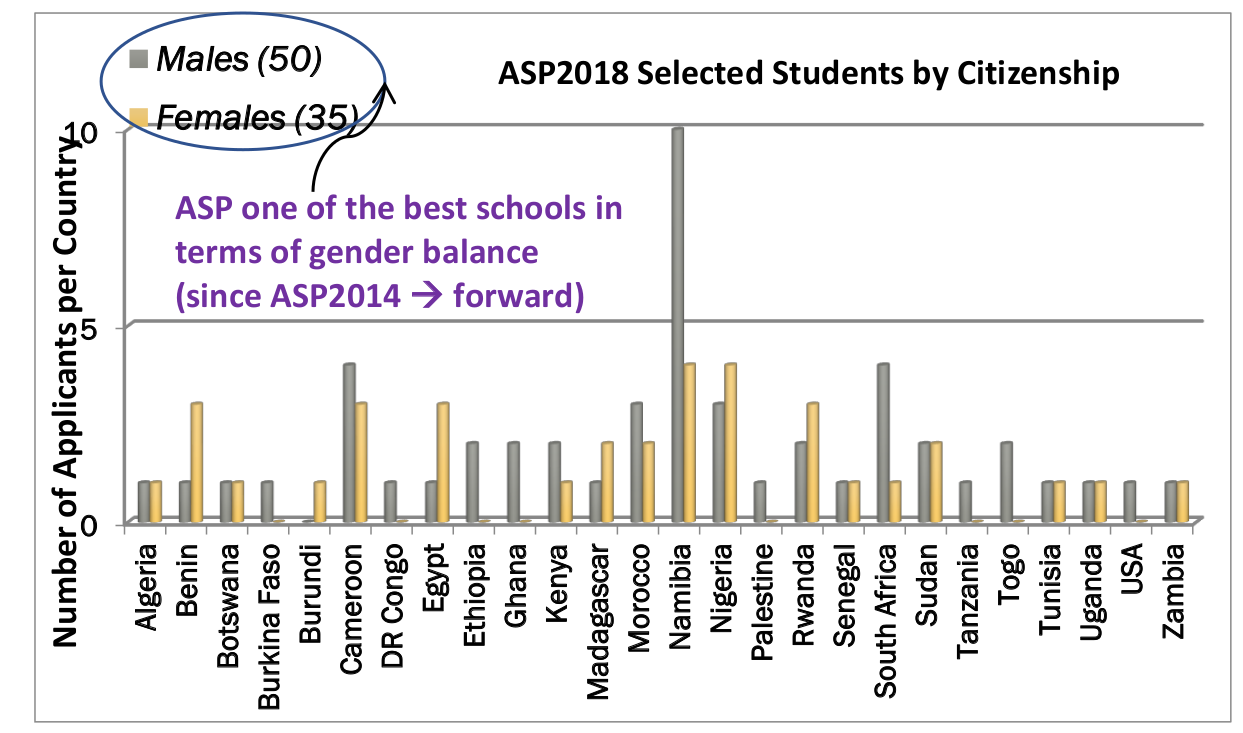}
\caption{Distribution of ASP2018 selected students by citizenship. There were 523 applicants from whom 85 were selected. There were an additional 30 good students on the waiting list. The selection is constrained by budget and logistical, and all the early declinations are replaced from the waiting list.}
\label{fig:1}
\end{figure}

\begin{figure}[!htp]
\centering
\includegraphics[height=2.8in]{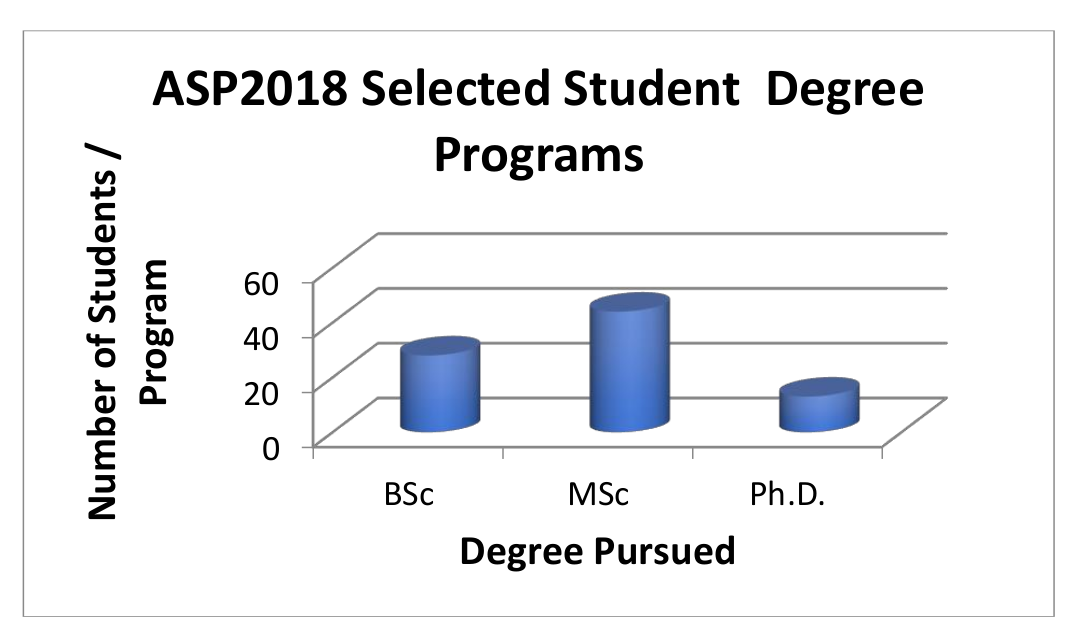}
\caption{The academic levels of the ASP2018 students at the time of the selections. Participating students mostly at the MSc level.}
\label{fig:2}
\end{figure}

\begin{figure}[!htp]
\centering
\includegraphics[height=3.1in]{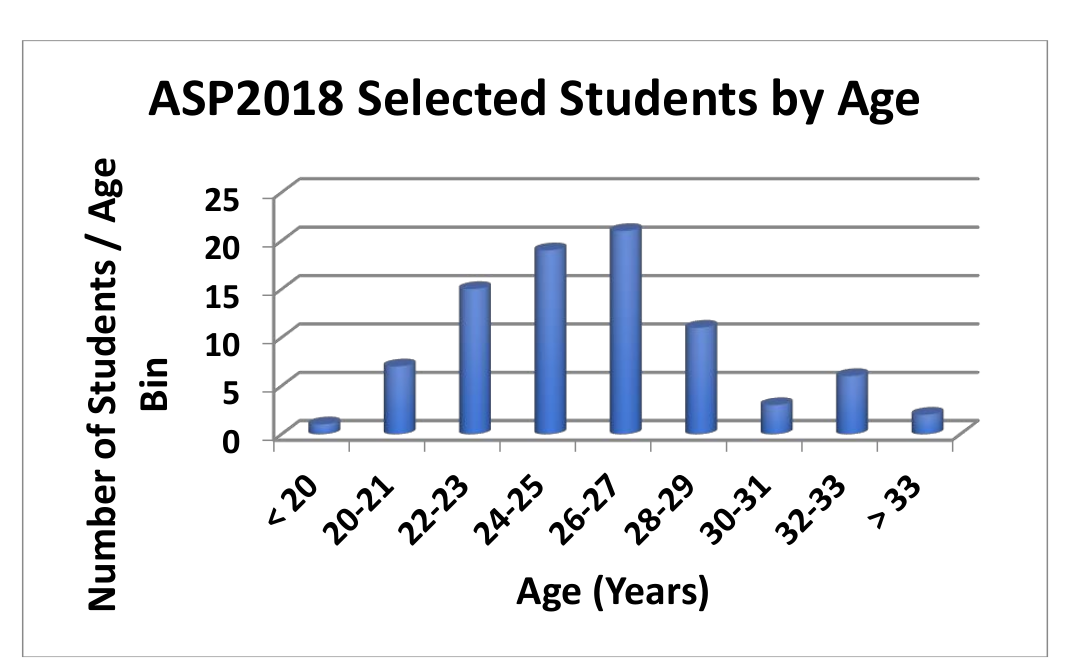}
\caption{The age distribution of the selected students. The peak is around 26-27 years old with the academic level at the MSc.}
\label{fig:3}
\end{figure}

\begin{figure}[!htp]
\centering
\includegraphics[height=4.1in]{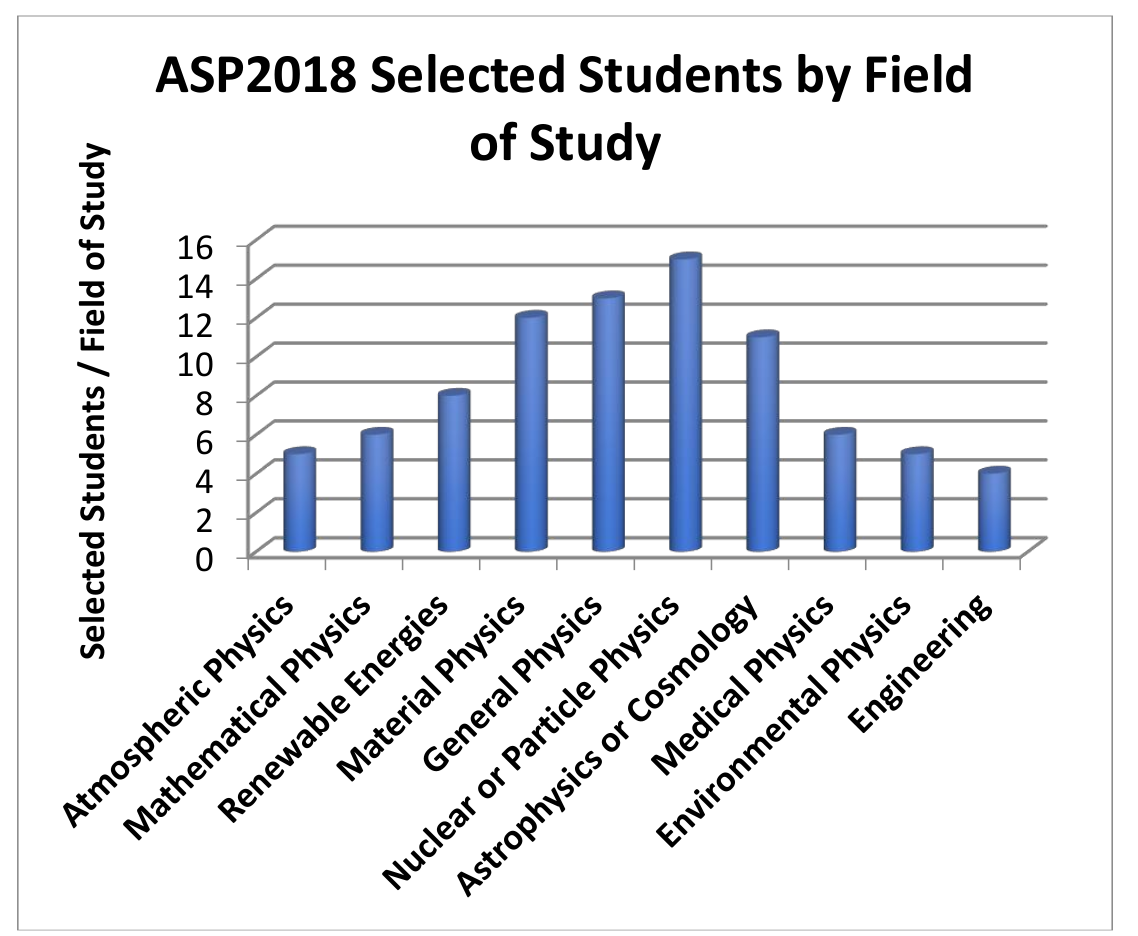}
\caption{The field of study of the ASP2018 students at the time of their selection. Fifteen to twenty percent of the selected students have their majors in particle physics. }
\label{fig:4}
\end{figure}

The school is biennially in different Africa, taking advantage of local support and considering a uniform exposure for Africa. The proposed duration of the school is three weeks. The host country of the school is selected two and a half years in advance through a competitive bidding process that considers aspects of safety of the participants and the support from the host country government. The target—in each edition of the ASP—is to have eight-five students selected from five hundred applicants in various African countries, seventy high school teachers from the host country for a one-week teachers training workshop, one thousand five hundred high school pupils for a one-week outreach, and sixty extra participants at the ASP conference. The objective of the ASP conference is to attract the participation of ASP alumni and African research faculties to give scientific talks and network with the international participants to foster new research collaborations. Peer-reviewed conference proceedings are published in the African Review of Physics. Full bursaries are provided to the selected students. \Cref{fig:1,fig:2,fig:3,fig:4} show the distributions of the selected students as a function of their citizenships, age and field of study. Alumni are not re-selected for the current school but they may attend the conference with the support of their academic institutes.

\section{Coordination of the activities}
ASP activities are coordinated by the International Organizing Committee (IOC) in collaboration with a Local Organizing Committee (LOC) in the host country. The IOC is advised by an International Advisory Committee (IAC). International lecturers (IL) help design the scientific program, which is often adapted to reflect the research interests of the host country. The lecturers also help in the selection and mentorship of the students. The ASP mentorship program is a dedicated effort to mentor ASP students in collaboration with their academic advisors; the mentorship program runs continuously even when there is no formal school. 

In the addition to the coordination of all the activities, the IOC is responsible for fundraising, identification of the next host country, and the activity reports to the funding agencies. Each funding agency has up to two representatives in the IAC to advise on the usage of the funds, the selection of the next host country, and on the rest of the activities. The LOC helps with all the local logistics and liaises with the education and research branches of the host country government.

\newpage

\section{Impact and retention}

\begin{figure}[!htp]
\centering
\includegraphics[height=2.8in]{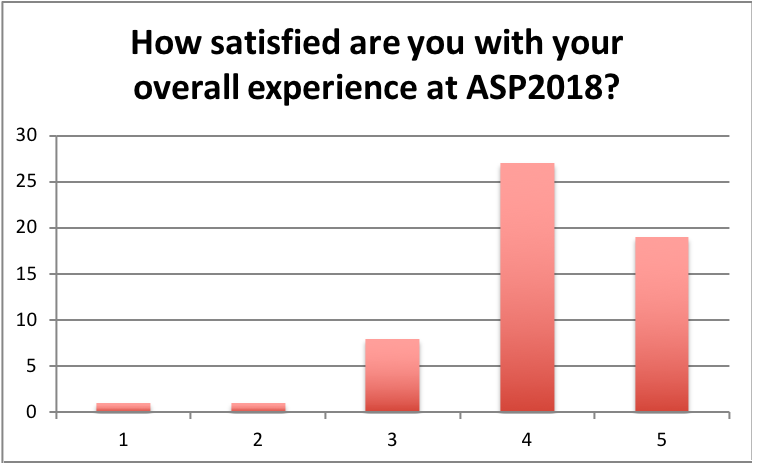}
\caption{students' satisfaction about their experience at ASP2018. "1" is the least satisfied and "5" shows the most satisfaction. }
\label{fig:5}
\end{figure}

\begin{figure}[!htp]
\centering
\includegraphics[height=3.3in]{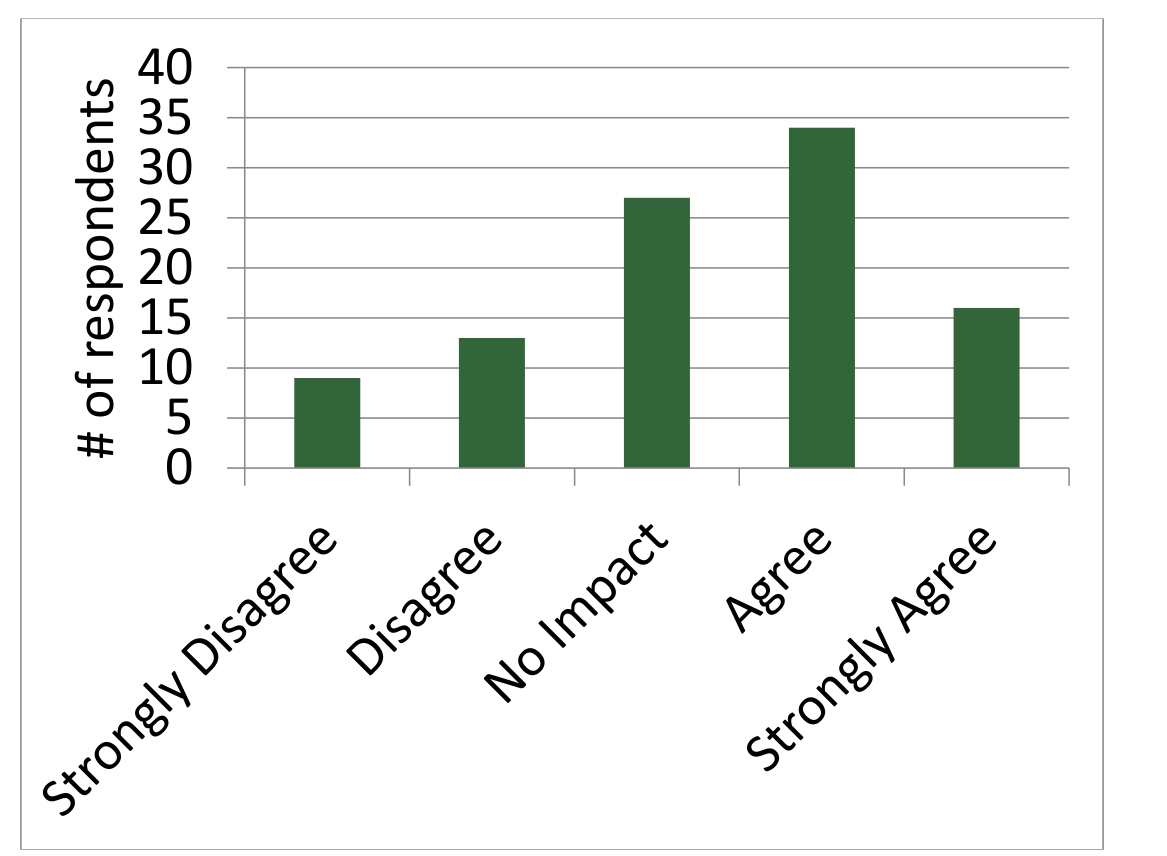}
\caption{The alumni were asked if their research has benefited from professional contacts made at ASP—during the school or afterward through mentorship. }
\label{fig:6}
\end{figure}

\begin{figure}[!htp]
\centering
\includegraphics[height=2.7in]{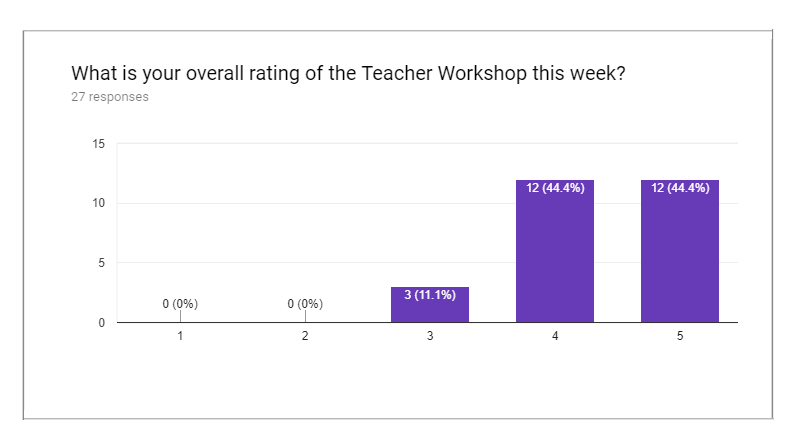}
\caption{ Feedback from the high school teachers on their experience during the workshop for teachers at the ASP2018 in Namibia.}
\label{fig:7}
\end{figure}

Two types of surveys are done about ASP. First, toward the end of the current edition of the school, the participating students and teachers are asked for feedback on their experience. The second survey is conducted every four years on all the alumni—this allows the IOC to follow the academic development of the alumni and answer questions such as, "What happens to the students after they have attended ASP?" or "Where are the students now?" The continuous mentorship program also helps answer these questions since the mentors follow the academic development of the students.  \Cref{fig:5,fig:6,fig:7} show some feedback from the students and high school teachers.

\section{Outlook}

The success of ASP is also due to the dedication of the international lecturers, most of whom cover their travels from external sources. This allows the funds raised by the IOC to be used to maximize the attendance of African students. Although the school duration is three weeks, most lecturers stay for only one of the three weeks on average to share their experience with African students. Some of the lecturers identify and help students pursue high education or research at their institutes.

One day during the school, the IOC organizes a forum on capacity development and retention in Africa. This forum provides the platform to engage African policymakers on physics education and research, the media, the industries and the international community in a discussion about the role of the ASP. The main objective of the forum is to improve the ASP program to serve the education and research priorities of African countries. 

\newpage
The ASP forum also encourages the promotion of research collaboration and consortia, e.g. ASP served as the platform for earlier discussions on the African Light Source ~\cite{connell2018african}. Through the ASP forum, collaboration among African countries are encouraged to share experience and approaches, and find concerted solutions to common issues in education.

The issues of retention—within Africa—of the qualified African graduates are also debated during the forum. The survey results show that most of ASP alumni stay in Africa—only 20$\%$ go outside for high education and research. The objective of the discussions on retention is to develop a mechanism to reintegrate into the Africa the alumni that go abroad.

\section{Conclusion}
We propose a school on fundamental physics and applications every two years in different African countries. The aim of the school would be to build capacity to harvest, interpret, and exploit the results of current and future physics experiments and to increase proficiency in related applications in Africa. The host countries are selected through competitive bids that address issues of safety of the participants and support from the local government. The school is opened to about 85 students from African countries selected by a committee of international lecturers, taking into account the need to promote fundamental physics and applications in developing countries. The school also includes a workshop to train high school teachers and an outreach for high school pupils from the host countries. An international conference is also added to the school activities to create a scientific networking environment between African participants and international delegates. Finally, a mentorship program running continuously even when there is no formal school allows African students to stay connected to lecturers from abroad and benefit from active mentorship to supplement the efforts of academic advisors.

\section*{Acknowledgements}
We would like to thank the universities, research institutes and government agencies that provided the funding for ASP schools.

Thanks also to the lecturers for the courses prepared and taught at the ASP schools, for the clarity of the materials presented to the students and for their availability during discussion and practical sessions to interact further with the students.

Finally, we thank each member of the organising committee (local and international) for responding to the challenge to prepare this school, for their concerted efforts to contribute to education in Africa. These were extra efforts beyond their professional obligations.


\begin{thebibliography}{99}


\bibitem{ASP}
K.A. Assamagan, B. Acharya, A.E. Dabrowski, C. Darve, J.R. Ellis, F. Ferroni, S.G. Muanza.  {\it The African School of Fundamental Physics and Applications}, https://www.africanschoolofphysics.org/.

\bibitem{ASP2010}
K. A. Assamagan et al., {\it African School of Fundamental Physics and its Applications, August 1-21, 2010, Stellenbosch, South Africa, \bf {ASP2010 Final Report}}, http://africanschoolofphysics.web.cern.ch/AfricanSchoolofPhysics/asp2010.pdf, December 2010.

\bibitem{ASP2012}
K. A. Assamagan et al., {\it African School of Fundamental Physics and its Applications, July 15 - August 8, 2012, Kumasi, Ghana, \bf {ASP2012 Final Report}}, http://africanschoolofphysics.web.cern.ch/AfricanSchoolofPhysics/asp2012/asp2012 final.pdf, Novemberr 2012.

\bibitem{ASP2014}
K. A. Assamagan et al., {\it African School of Fundamental Physics and its Applications, August 3-23 2014, Dakar, Senegal, \bf {ASP2014 Final Report}}, http://www.africanschoolofphysics.org/wp-content/uploads/2014/11/asp2014.pdf,
November 2014.

\bibitem{ASP2016}
K. A. Assamagan et al., {\it African School of Fundamental Physics and its Applications, August 1-19 2016, Kigali, Rwanda, \bf {ASP2016 Final Report}}, https://ketevi.web.cern.ch/ketevi/ASP2016/asp2016-FinalReport.pdf, September
2016.

\bibitem{ASP2018}
K. A. Assamagan et al., {\it African School of Fundamental Physics and its Applications, June 24-July 14 2018, Windhoek, Namibia, \bf {ASP2018 Final Report}}, https://ketevi.web.cern.ch/ketevi/ASP2018/FinalReport/asp2018.pdf, September 2018

\bibitem{connell2018african}
S.H. Connell et. al.,  {\it The African light source project}, Afr Rev Phys, {\bf 13} 0019 (2018).


\end{thebibliography}
\end{document}